\documentclass[aps,prb,twocolumn]{revtex4-1}
\usepackage{graphicx}

\begin{document}

%%%%%%%%%%%%%%%%%%%%%%%%%%%%%%%%%%%%%%%%%%%%%%%%%%%%%%%%%%%%%%%%%%%%%%%%%%%%%%%%%%%%%%%%%%%%%%%%%%%%%%%%%%%%%%%%%%%%
\title{Creating new layered structures at high pressures: SiS$_2$}

%%%%%%%%%%%%%%%%%%%%%%%%%%%%%%%%%%%%%%%%%%%%%%%%%%%%%%%%%%%%%%%%%%%%%%%%%%%%%%%%%%%%%%%%%%%%%%%%%%%%%%%%%%%%%%%%%%%%
\author{Du\v{s}an Pla\v{s}ienka}
\email{plasienka@fmph.uniba.sk}
\affiliation{Department of Experimental Physics, Comenius University, Mlynsk\'{a} Dolina F2, 842 48 Bratislava, Slovakia}
\author{Roman Marto\v{n}\'{a}k}
\affiliation{Department of Experimental Physics, Comenius University, Mlynsk\'{a} Dolina F2, 842 48 Bratislava, Slovakia}
\author{Erio Tosatti}
\affiliation{International School for Advanced Studies (SISSA) and CNR-IOM Democritos, Via Bonomea 265, 34136 Trieste, Italy}
\affiliation{The Abdus Salam International Centre for Theoretical Physics (ICTP), Strada Costira 11, 34151 Trieste, Italy}

\date{\today}

%%%%%%%%%%%%%%%%%%%%%%%%%%%%%%%%%%%%%%%%%%%%%%%%%%%%%%%%%%%%%%%%%%%%%%%%%%%%%%%%%%%%%%%%%%%%%%%%%%%%%%%%%%%%%%%%%%%%
\begin{abstract}

Old and novel layered structures are attracting increasing attention for their physical, electronic, and frictional properties. SiS$_2$, isoelectronic to SiO$_2$, CO$_2$ and CS$_2$, is 
a material whose phases known experimentally up to 6 GPa exhibit 1D chain-like, 2D layered and 3D tetrahedral structures. We present highly predictive $ab$ $initio$ calculations combined with 
evolutionary structure search and molecular dynamics simulations of the structural and electronic evolution of SiS$_2$ up to 100 GPa. A highly stable CdI$_2$-type layered structure, which
is octahedrally coordinated with space group $P\bar{3}m1$ surprisingly appears between 4 and up to at least 100 GPa. The tetrahedral-octahedral switch is naturally expected upon compression,
unlike the layered character realized here by edge-sharing SiS$_6$ octahedral units connecting within but not among sheets. The predicted phase is semiconducting with an indirect band gap of 
about 2 eV at 10 GPa, decreasing under pressure until metallization around 40 GPa. The robustness of the layered phase suggests possible recovery at ambient pressure, where calculated
phonon spectra indicate dynamical stability. Even a single monolayer is found to be dynamically stable in isolation, suggesting that it could possibly be sheared or exfoliated from
bulk $P\bar{3}m1$-SiS$_2$.

\vspace{1cm}

\textbf{This work has been submitted into the \textit{Scientific Reports} journal}

\end{abstract}

%%%%%%%%%%%%%%%%%%%%%%%%%%%%%%%%%%%%%%%%%%%%%%%%%%%%%%%%%%%%%%%%%%%%%%%%%%%%%%%%%%%%%%%%%%%%%%%%%%%%%%%%%%%%%%%%%%%%
\maketitle
%%%%%%%%%%%%%%%%%%%%%%%%%%%%%%%%%%%%%%%%%%%%%%%%%%%%%%%%%%%%%%%%%%%%%%%%%%%%%%%%%%%%%%%%%%%%%%%%%%%%%%%%%%%%%%%%%%%%

%%%%%%%%%%%%%%%%%%%%%%%%%%%%%%%%%%%%%%%%%%%%%%%%%%%%%%%%%%%%%%%%%%%%%%%%%%%%%%%%%%%%%%%%%%%%%%%%%%%%%%%%%%%%%%%%%%%%
\section*{Introduction}

SiS$_2$ is a member of an important family of group IV-VI AB$_2$ compounds made of light elements including well-known
materials such as CO$_2$ \cite{Santoro-Gorelli-CO2-review, Yoo-CO2-review-1, Iota-Yoo-Cynn, Yoo-1, Datchi-V, Santoro-3, Iota-Yoo, Yoo-2, Bonev, Sengupta-3, Tschauner, Togo, Serra, Sun},
SiO$_2$ \cite{Tse-SiO2, Dove, Murakami, Prakapenka-SiO2, Dera, Nekrashevich-Gritsenko, cell-metadynamics-3, Hu-SiO2},
GeO$_2$ \cite{Haines, Lodziana, Prakapenka-GeO2, Ono, Micolaut-GeO2-review}
and CS$_2$ \cite{Yuan-Ding, Dias, Zarifi, Naghavi}. Limited to relatively low pressures, the structural evolution of SiS$_2$ has also been accurately described. \cite{Evers}
The ambient-pressure stable phase  known as NP-SiS$_2$ has orthorhombic $Ibam$ structure and consists of distorted edge-sharing tetrahedra 
forming 1D chains which interact via weak van der Waals forces \cite{Silverman-Soulen, Prewitt-Young, Peters-Krebs, Guseva, Zwijnenburg, Bletskan, Tokuda, Evers, Wang-SiSe2}.
At 2.8 GPa, a first high-pressure phase HP1-SiS$_2$ \cite{Guseva, Evers} appears, with monoclinic space group $P2_1/c$, by interconnection of the NP-SiS$_2$ chains to form 2D layers
\cite{Evers}. The connectivity pattern changes from each tetrahedron sharing two edges in the NP phase to sharing one edge and two corners in HP1 \cite{Evers}. 
The very same structure was recently predicted to be stable at pressures beyond 30 GPa in CS$_2$ \cite{Naghavi}. Further increase of pressure leads to the 
HP2-SiS$_2$ phase at 3.5 GPa \cite{Guseva, Evers}, again with $P2_1/c$ space group, and again with edge- and corner-sharing tetrahedra, 
in this case, however, with a 3D connectivity network with large cavities \cite{Evers}. Finally, at 4 GPa, a tetragonal $I\bar{4}2d$ structure denoted as HP3-SiS$_2$ takes over,
a phase formed by strictly corner-sharing tetrahedra that are slightly distorted and span the whole three-dimensional space \cite{Silverman-Soulen, Prewitt-Young, Guseva, Zwijnenburg, Bletskan, Evers}.
The same structure as HP3-SiS$_2$ is adopted by CO$_2$ phase V at high pressures \cite{Datchi-V} and it can be viewed as a partially collapsed version of SiO$_2$ $\beta$-cristobalite \cite{Santoro-3}.

The phase diagram of silicon disulfide at pressures higher than 6 GPa remains unknown experimentally. Evers $et$ $al.$ \cite{Evers} reasonably hypothesized that 
at higher pressures such as 10-20 GPa, SiS$_2$ might adopt a six-fold, octahedral Si coordination, accompanying a density increase. An alternative scenario might be the creation 
of a denser tetrahedral phase such as coesite in SiO$_2$. Work is clearly needed to explore the high pressure phases of SiS$_2$. While of course future experimental work
is called for, experience of the last decade has shown that state-of-the-art density-functional based crystal structure search can be extremely predictive. 

By means of a well-tested, highly reliable protocol consisting of  $ab$ $initio$ electronic structure calculations combined with evolutionary search for crystal structure
prediction \cite{Oganov-Glass-USPEX, XtalOpt} and with constant-pressure molecular dynamics (MD) simulations, we undertook a fresh theoretical exploration of high-pressure phases of 
SiS$_2$ up to 100 GPa. While confirming first of all the presence and stability of the known low-pressure tetrahedral phases, our study predicts three new structures with octahedral, six-fold  
Si coordination at higher pressures. While that confirms previous expectations, the surprise is that the new structures are layered. We analyze their structural and electronic properties 
in detail at all pressures, including properties of a single octahedral monolayer which is predicted to survive in a metastable state, should one succeed to shear off and exfoliate it away from the
bulk layered structure. More generally, these findings suggest the idea to create, shear and possibly stabilize one or more SiS$_2$ semiconducting monolayers with over 2 eV band gap on a
substrate in some future realizations.

%%%%%%%%%%%%%%%%%%%%%%%%%%%%%%%%%%%%%%%%%%%%%%%%%%%%%%%%%%%%%%%%%%%%%%%%%%%%%%%%%%%%%%%%%%%%%%%%%%%%%%%%%%%%%%%%%%%%
\section*{Results}

\begin{figure*}
\centering
\includegraphics[width=2.0\columnwidth]{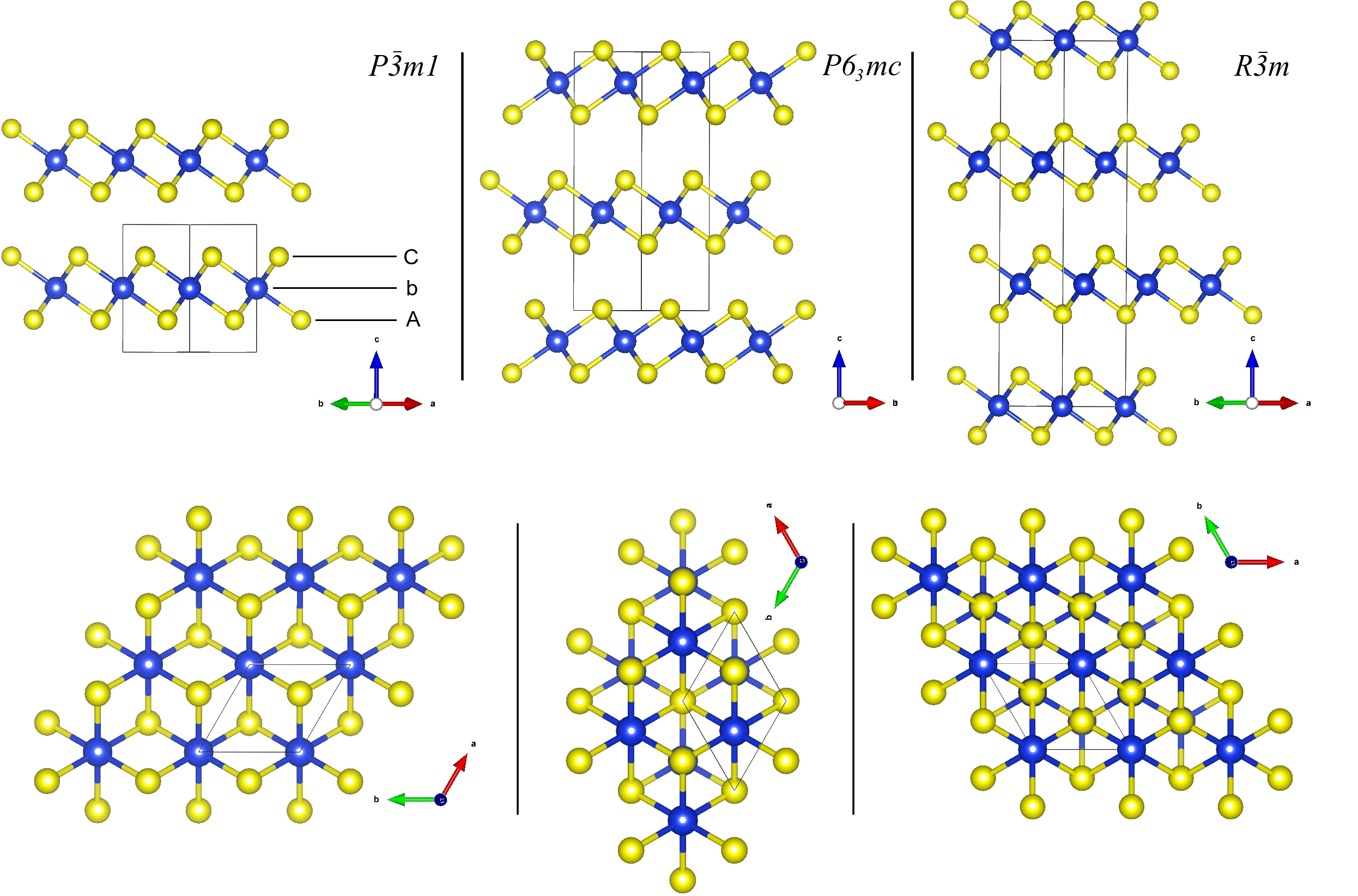}
\caption{$2\times 2 \times 1$ supercell of $P\bar{3}m1$ structure with single sheet per period (left pictures), $P6_3mc$ with two sheets
(middle pictures) and $R\bar{3}m$ with three sheets per unit cell (right pictures). Upper pictures show different stackings along the lateral direction parallel to planes,
while lower pictures are views from direction axial to the sheets, which corresponds to $c$-axis vector view in all three phases.
Black boxes represent the respective primitive cells with one, two and three SiS$_2$ units for $P\bar{3}m1$, $P6_3mc$ and $R\bar{3}m$ phases, respectively.
Along the axial view, sheets in $P\bar{3}m1$ overlap.}
\label{all}
\end{figure*}

%%%%%%%%%%%%%%%%%%%%%%%%%%%%%%%% structures %%%%%%%%%%%%%%%%%%%%%%%%%%%%%%%%%%%
\subsection{Phases of SiS$_2$ above the tetrahedral regime}

From the extensive evolutionary search up to 100 GPa, MD simulations and structural optimizations (details given in Methods)  we found three most
promising candidates for high-pressure phases of SiS$_2$ beyond 6 GPa. % currently the highest pressure at which SiS$_2$ was determined experimentally.
These three lowest-enthalpy phases have space groups $P\bar{3}m1$, $P6_3mc$ and $R\bar{3}m$ - Fig. \ref{all} and are very similar to each other. They all consistently emerged at every 
investigated pressure (10, 30, 60 and 100 GPa) in the evolutionary search. All three are layered, formed by separate sheets of edge-sharing SiS$_6$ octahedra
with different stacking - Fig. \ref{all}. The layered character, which is of potential interest in view of a possibly facile frictional sliding under high-pressure shear,
is remarkable, since in most materials one tends to associate higher density with ''more 3D'' networks - precisely the trend observed in the SiS$_2$ tetrahedral 
region (from NP to HP1, HP2 and HP3 between 0 and 4 GPa).

\begin{figure}
\centering
\includegraphics[width=\columnwidth]{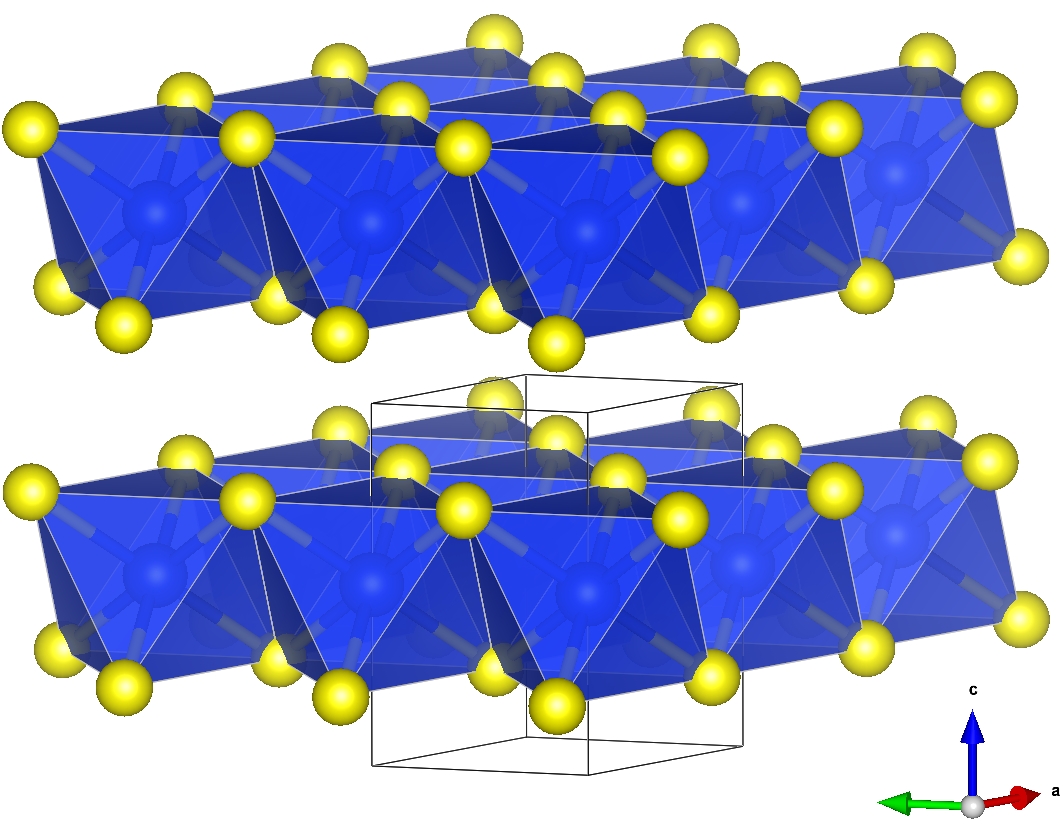}
\caption{The most stable and the most dense $P\bar{3}m1$ phase (C6 structure) of SiS$_2$ with one sheet per unit cell.
Polyhedral view visualized by VESTA \cite{VESTA} shows interbonded edge-sharing octahedra forming individual sheets.}
\label{P-3m1}
\end{figure}

In these novel layered phases each of the octahedra shares six out of its twelve edges with six surrounding octahedra. The most stable phase is $P\bar{3}m1$ -
shown in Fig. \ref{P-3m1} and in Fig. \ref{all} (left pictures), which contains only one SiS$_2$ unit in the primitive cell. It is isostructural to
CdI$_2$ (C6 structure) \cite{Wilson-Yoffe} where sheets are stacked directly above each other. The same structure is found at ambient pressure in chemically similar
systems with heavier atoms such as SnS$_2$ \cite{Burton-SnS2} or SiTe$_2$ \cite{Mishra-SiTe2} as well as in several other chalcogenides, iodides, chlorides and
bromides \cite{Wilson-Yoffe, Greenwood-CdI2, Ang, Aksoy-TiS2, Bao-TiS2}, or even in BeH$_2$ \cite{Wang-BeH2}.
The second lowest-enthalpy phase $P6_3mc$ contains two layers per unit cell which are mutually staggered and reflected with respect to each other - Fig. \ref{all} (middle pictures).
Finally, the least stable $R\bar{3}m$ phase shown in Fig. \ref{all} (right pictures) contains three sheets per unit cell and its structure is identical to ambient-conditions CdCl$_2$.

The stacking pattern of the structures is illustrated in Fig. \ref{all} by lateral (upper pictures) and axial (lower pictures) views with respect to sheets.
The arrangement of sheets in all three phases may be represented by specifying the stacking pattern of individual single-atomic layers of Si and S atoms,
as far as these form planes of closed-packed atoms that are placed regularly one on top of another. The stacking pattern of the $P\bar{3}m1$ structure
in this single-atom layer notation is $/AbC/$ \cite{Wilson-Yoffe}, while for $P6_3mc$ it is $/AbC/AcB/$ and for $R\bar{3}m$ it is $/AbC/BcA/CaB/$, where small letters
denote layers of Si atoms and capital letters are for planes of S atoms (see legend in Fig. \ref{all} lateral view on $P\bar{3}m1$).
Structural data of the three phases at 10 GPa, their densities and band gaps calculated with generalized-gradient approximation (GGA) are summarized in Table \ref{data}.

%%%%%%%%%%%%%%%%%%%%%%%%%%%%%%%%%%%%%%%%%% structural data %%%%%%%%%%%%%%%%%%%%%%%%%%%%%%%%%%%%%%%%%%%%%%
\begin{table*}
\begin{tabular}{|l|l|c|l|c|c|c|}

\hline
phase symmetry    & unit-cell parameters & $Z$  & Wyckoff positions &    density     & PBE band gap \\
 (prototype)      &        [\AA]         &                                                         &                   & [g.cm$^{-3}$]  &    [eV]     \\ \hline

$P\bar{3}m1$ ($\#164$)       & $a=$ 3.213    &  1  &   Si1  \ \, \, 1b \, \, \,  0  \, \, \,  0  \, \, \, 1/2     &  3.226   &  0.9   \\
trigonal (CdI$_2$-type) & $c=$ 5.310    &     &    S1  \, \, \, 2d \, \, \, 1/3 \, 2/3 \, 0.2521  &          &        \\ \hline

$P6_3mc$ ($\#186$) & $a=$ 3.213         &  2  &   Si1  \ \, \, 2b \, \, \, 1/3 \, 2/3 \, 0.8799  &  3.223  & 0.7 \\
hexagonal          & $c=$ 10.629        &     &    S1  \, \, \,\ 2b \, \, \, 2/3 \, 1/3 \, 0.0035  &         &     \\
                   &                    &     &    S3  \, \, \,\ 2a \, \, \,  0  \, \, \, 0  \, \, \, 0.2559  &         &     \\ \hline

$R\bar{3}m$ ($\#166$)         & $a=$ 3.213   &  3  &   Si1  \ \, \, 3a \, \, \, 2/3 \, 1/3 \, 1/3     &  3.201  &  0.5     \\
trigonal (CdCl$_2$-type) & $c=$ 16.052  &     &    S1  \ \, \, \, 6c \, \, \, 1/3 \, 2/3 \, 0.4151  &         &     \\ \hline

\end{tabular}
\label{data}
\caption*{Structural data, density and PBE band gap of the three proposed layered structures of SiS$_2$ at 10 GPa.}
\end{table*}

%%%%%%%%%%%%%%%%%%%%%%%%%%%%%%%%%%%%%%%%%%%%%%%%%%%%%%%%%%%%%%%%%%%%%%%%%%%%%%%%%%%%%%%%%%%%%%%%%%%%%%%%%%%%%%%%%%%%
\subsection*{Stability of high-pressure octahedral layered structures}

\begin{figure}
\centering
\begin{tabular}{cc}
\includegraphics[width=\columnwidth]{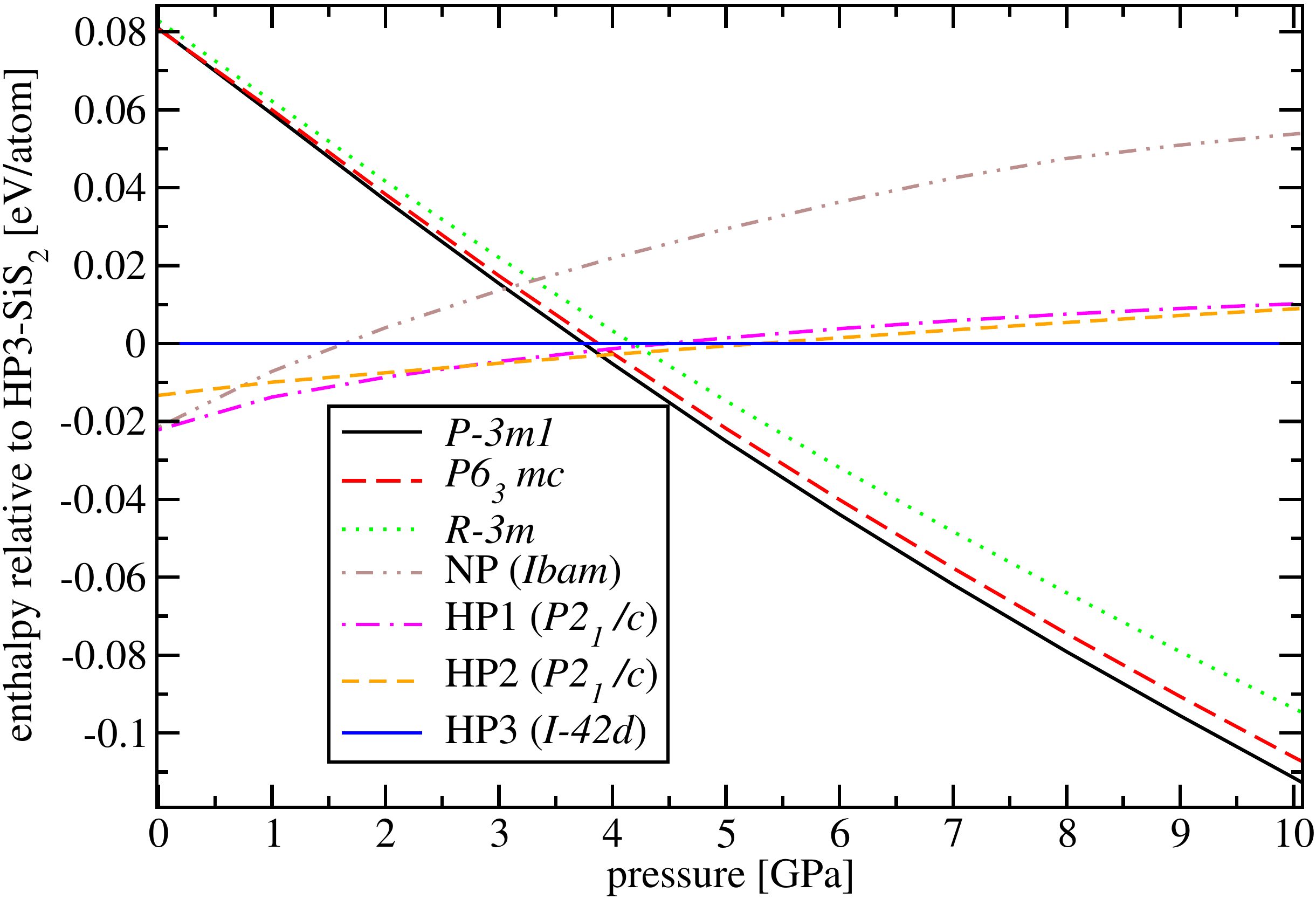}\\
\includegraphics[width=\columnwidth]{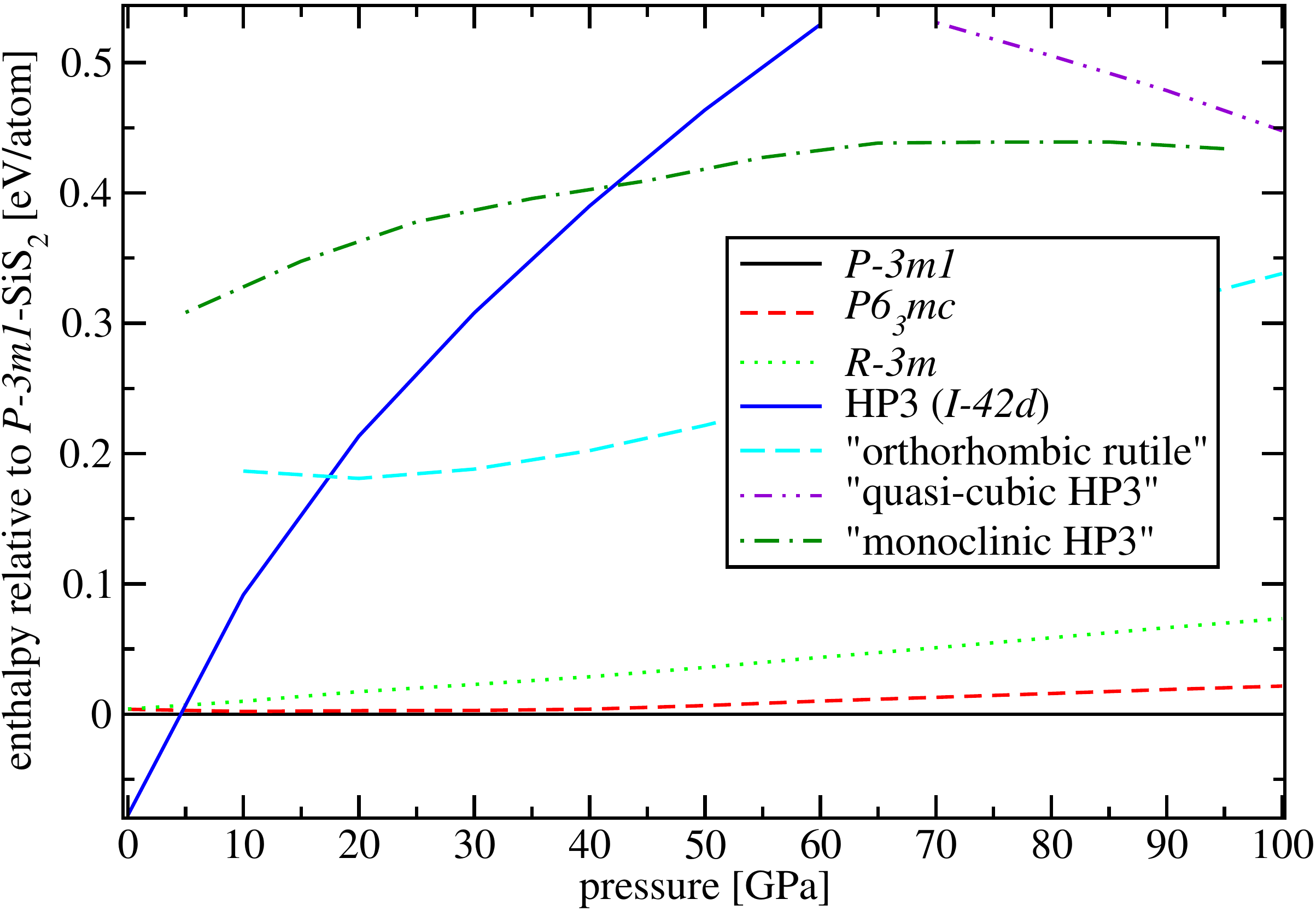}\\
\end{tabular}
\caption{(Upper panel) Enthalpies of octahedral structures $P\bar{3}m1$ (black solid curve), $P6_3mc$ (red dashed curve) and $R\bar{3}m$
(green dotted curve) and of tetrahedral NP (brown dashed-double dotted), HP1 (violet dashed-dotted) and HP2 (orange dashed) forms relative to HP3 (blue solid horizontal line)
calculated up to 10 GPa. Layered octahedral phases become more stable than tetrahedral forms over 4 GPa.
(Lower panel) Enthalpies relative to $P\bar{3}m1$ of the three octahedral phases, HP3 and of three newly identified metastable forms of SiS$_2$
(orthorhombic version of rutile structure, quasi-cubic and monoclinic versions of HP3) in 1 Mbar range showing a strong tendency towards forming octahedral forms.}
\label{enthalpies}
\end{figure}

Calculated $T$=0 enthalpies show that the sequence of transitions for tetrahedral phases occurring at 3.5 GPa (HP1$\rightarrow$HP2) and 4 GPa (HP2$\rightarrow$HP3)
experimentally \cite{Evers} at room temperature is reasonably well reproduced within our calculations, which predict these two transitions to take place
(at zero temperature) at 2.8 and 5.4 GPa, respectively - Fig. \ref{enthalpies} (upper panel). However,
all three six-coordinated phases become more stable than HP2-SiS$_2$ above cca. 4 GPa. Hence, as far as HP3 crosses the HP2 curve only at 5.4 GPa, the calculations
predict that the HP3 phase should be metastable at $T$ = 0 at all pressures. Its experimental observation between 4-6 GPa \cite{Evers} might be due to temperature, or to kinetics. 
The NP-HP1 transition occurred at 2.8 GPa experimentally \cite{Evers} while theoretically these two phases have practically the same enthalpy at $p$=0 while that of NP rises considerably 
faster with pressure. These discrepancies between the experimental and theoretical phase boundaries, including the predicted metastability of the HP3 phase at zero temperature,
might have several causes. For example, one of the most probable reason for why octahedral geometry was not yet found in laboratory could be the effect of hysteresis that
is often exhibited by first-order phase transitions. The transition HP2(HP3)$\rightarrow P\bar{3}m1$ is indeed strongly first-order (see later), which implies possible large
energy barriers hindering the transformation between the two. In addition to kinetic effects, also thermal effect of vibrational entropy might play a role
here and perhaps it may explain the fact that normally only the NP phase is observed at ambient conditions. Moreover, inaccuracies of density functional-based methods 
as well as pressure measurements might also be the source of disagreements. Nevertheless, calculations showed that in the pressure range of 2.8 - 5.4 GPa there exist a number of structures
with very similar enthalpy and so accurate determination of equilibrium phase boundaries in this region might be complicated from both experimental and theoretical perspective.

More generally, even if this level of agreement between calculations and experiments below 6 GPa may seem somewhat imperfect, it is important to stress that it is on the contrary 
quite good, because at low pressures all structural enthalpy differences are generally small, and more dependent upon approximations. Both the predictive quality of calculations 
and the enthalpy differences increase with pressure, and that gives us confidence in our high-pressure study.

Strong structural similarity between octahedral phases implies similar thermodynamical properties. % especially at low pressures, as represented e.g. by enthalpies.
The enthalpy difference between the most stable $P\bar{3}m1$ and the least stable $R\bar{3}m$ layered octahedral phase is cca. 15 meV/atom at 10 GPa - Fig. \ref{enthalpies} (upper panel),
which indicates that in experiments one could possibly find SiS$_2$ in a state of irregularly stacked polytypes. This enthalpy difference is composed of an internal energy
difference of less than 5 meV/atom and $p\Delta V$ term of about 10 meV/atom. The energy differences remain similar at all pressures,
probably because both inter-sheet repulsion and strong intraplanar covalent bonding change with pressure in the same way in all phases. The overall enthalpy
difference, however, grows with pressure and reaches 73 meV/atom at 100 GPa - Fig. \ref{enthalpies} (lower panel). That can be attributed to the $pV$ term 
which favors phases with more efficient packing, the best being the simplest $P\bar{3}m1$ structure with one sheet per cell (see densities in Table \ref{data}).

During the evolutionary search, we identified numerous different non-octahedral phases of SiS$_2$, however, all of these were found to be grossly metastable at pressures below 100 GPa.
Most of these structures are tetrahedral and among them, the HP3 phase was found to possess the lowest enthalpy at lower pressures, yet at 10 GPa it is still about 90 meV/atom higher
in enthalpy compared to the three layered octahedral forms and this difference rises quickly with pressure - Fig. \ref{enthalpies} (lower panel).
Some of the identified metastable structures (the quasi-cubic and monoclinic versions of the HP3-SiS$_2$ phase and orthorhombic-like version of the rutile phase)
were included into the 0-100 GPa enthalpy graph to show strong preference of SiS$_2$ to form six-coordinated forms in the investigated pressure range.

At 4 GPa, the calculated density of octahedral $P\bar{3}m1$ is 3.077 g.cm$^{-3}$, while that calculated for tetrahedral HP2 is 2.530 g.cm$^{-3}$ and that of HP3 is 2.577 g.cm$^{-3}$.
The large density jump between $P\bar{3}m1$ and HP2 (HP3)  - 19.4\% (21.6\%)  shows that despite the layered character, that was not predicted, other expectation 
of an octahedral phase by Evers $et$ $al.$ \cite{Evers} are well borne out. The strong first-order character of this structural transition, the large enthalpy gain and
density jump suggest some more qualitative but interesting points. First, this transition (or a very similar) is absolutely inevitable in  SiS$_2$ under pressure.  Second, it will 
necessarily involve a large hysteresis, with many possible metastability phenomena en route. Third, once created the new phase will be protected by large free energy barriers.
With some qualitative analogy with graphite-diamond (where of course the density jump is a much larger 55\% ) these large barriers and the associated nucleation costs 
might permit to the layered phases to survive in a metastable state, once recovered at ambient pressure and low temperatures.

%%%%%%%%%%%%%%%%%%%%%%%%%%%%%%%%%%%%%%%%%%%%%%%%%%%%%%%%%%%%%%%%%%%%%%%%%%%%%%%%%%%%%%%%%%%%%%%%%%%%%%%%%%%%%%%%%%%%
\subsection*{Electronic structure}

\begin{figure}
\centering
\begin{tabular}{ccc}
\includegraphics[width=\columnwidth]{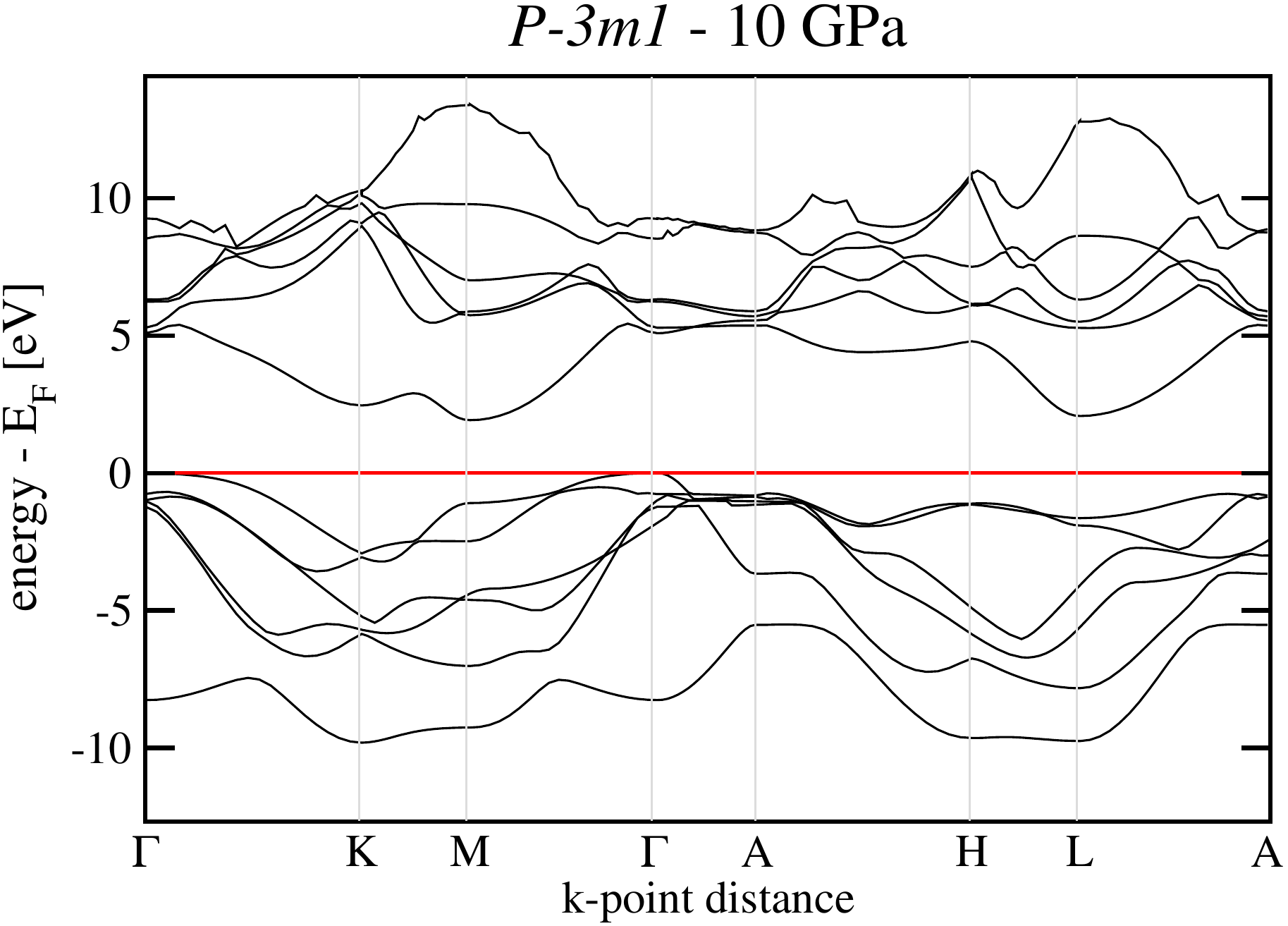}\\
\includegraphics[width=\columnwidth]{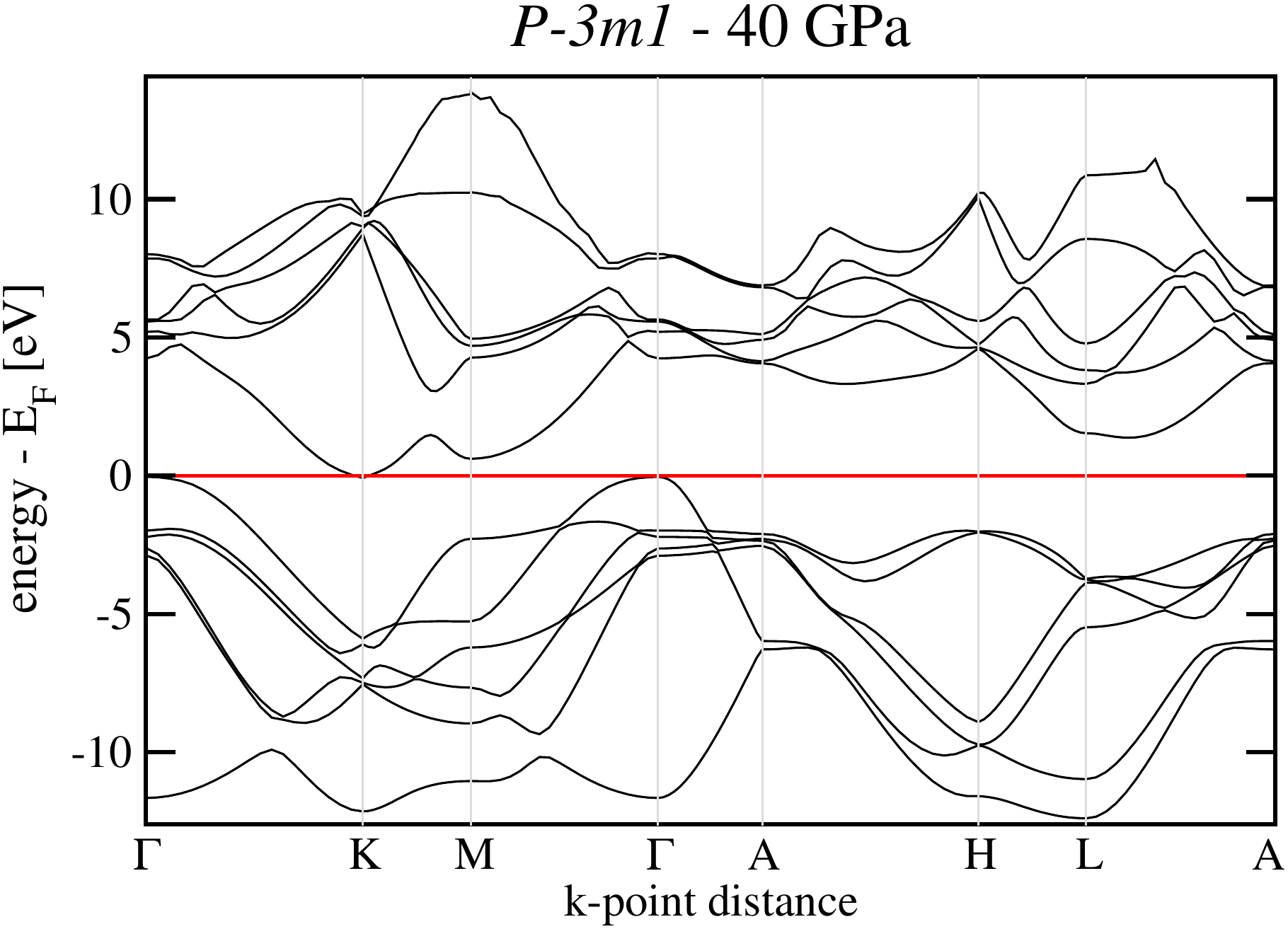}\\
\includegraphics[width=0.6\columnwidth]{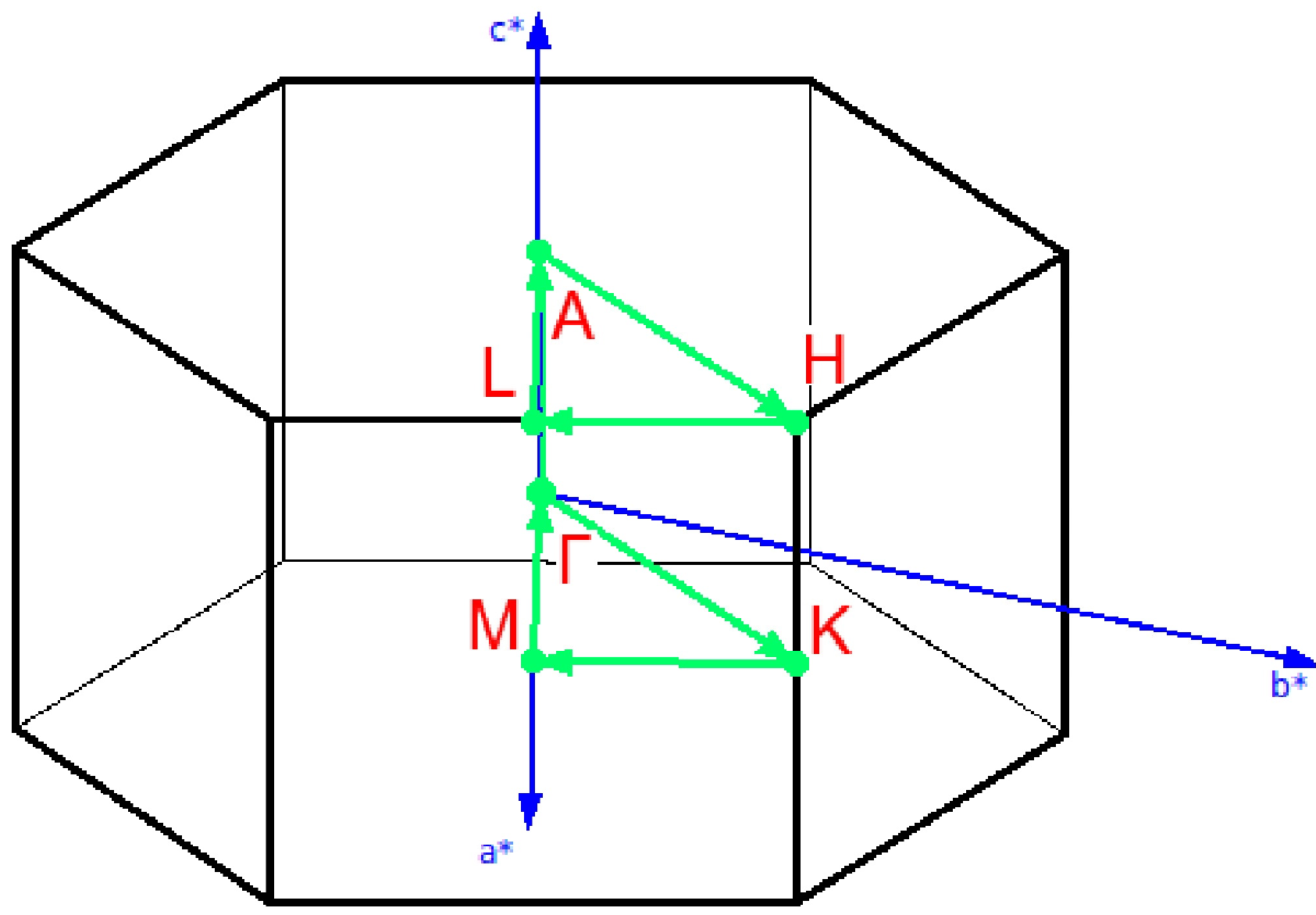}\\
\end{tabular}
\caption{HSE06 electronic band structure of $P\bar{3}m1$ in the semiconducting state at 10 GPa (upper panel) and at 40 GPa (middle panel), where the system begins to metallize.
In the lower panel, the $\Gamma \rightarrow K \rightarrow M \rightarrow \Gamma \rightarrow A \rightarrow H \rightarrow L \rightarrow A$ path in the first Brillouin zone is visualized.}
\label{bands}
\end{figure}

At their appearance above 4GPa all new octahedral layered phases are indirect gap semiconductors. The values of GGA band gaps at 10 GPa for the investigated phases, 
of order 2 eV, are given in Table \ref{data}. At 30 GPa, the gaps close and all layered forms are predicted to metallize.
However, it is a well-known fact that GGA approximation tends to underestimate the value of the band gap and the pressure of metallization
and therefore we repeated the electronic band structure calculations for $P\bar{3}m1$ also using the HSE06 hybrid functional that includes exact exchange.
Within HSE06, probably more reliable in this respect, all three layered octahedral phases metallize around 40 GPa. In Fig. \ref{bands}, the HSE06 electronic band structure
of $P\bar{3}m1$ is shown for the semiconducting state at 10 GPa (upper panel) and at the pressure of metallization at 40 GPa (middle panel),
along the corresponding $k$-space path. At 40 GPa the gap closes by band overlap between the $\Gamma$ point (top of the valence band, with large S-character) 
and the K point (bottom of the conduction band, with large Si-character).

\begin{figure}
\centering
\includegraphics[width=\columnwidth]{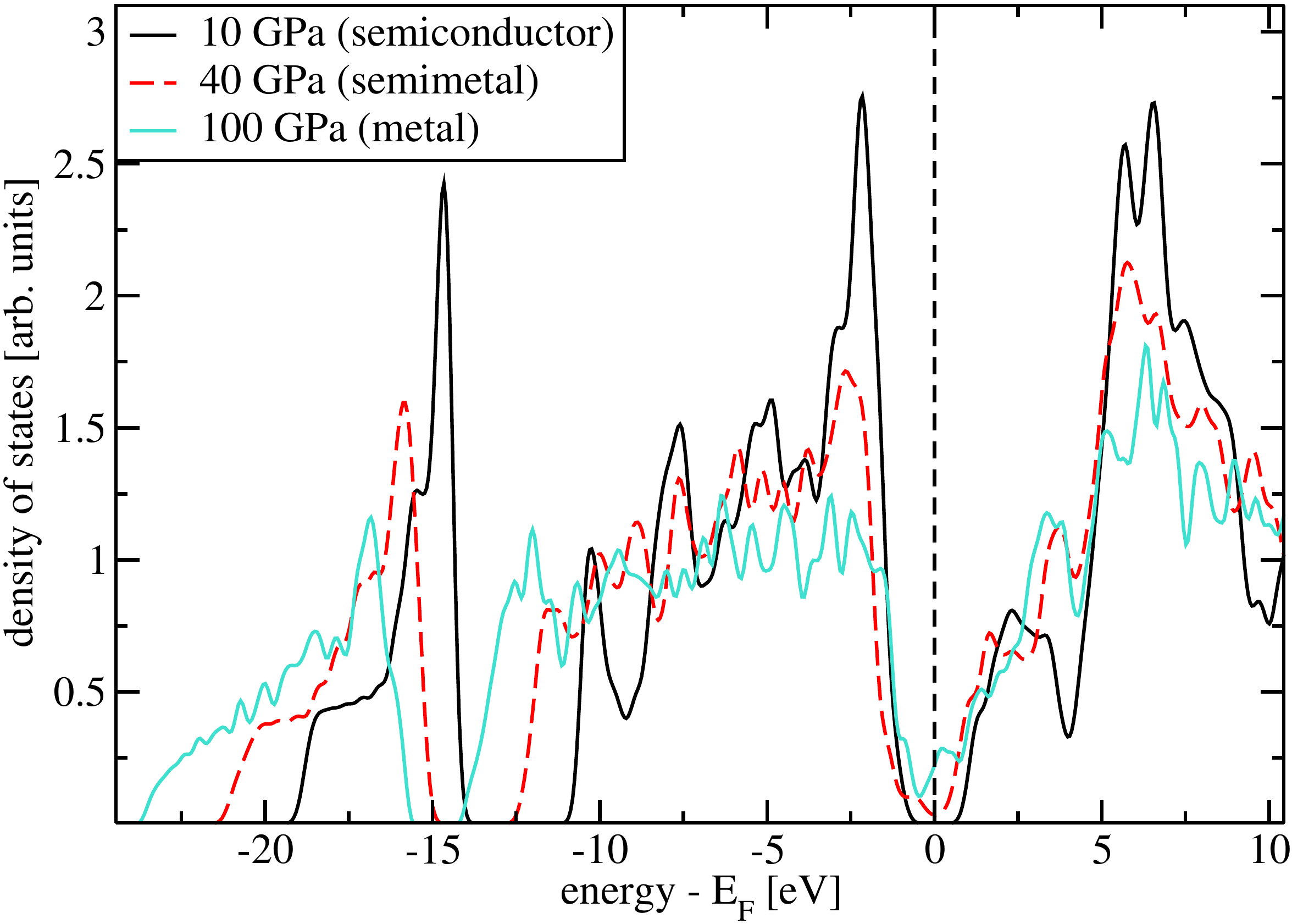}
\caption{HSE06 density of states of $P\bar{3}m1$ at 10 GPa, 40 GPa and at 100 GPa, where SiS$_2$ is already metallic.}
\label{dos}
\end{figure}

Electronic densities of states of $P\bar{3}m1$ at 10, 40 and at 100 GPa are shown in Fig. \ref{dos}. After metallization layered SiS$_2$ remains a rather poor metal 
with low density of states near the Fermi energy even at 100 GPa.

%%%%%%%%%%%%%%%%%%%%%%%%%%%%%%%%%%%%%%%%%%%%%%%%%%%%%%%%%%%%%%%%%%%%%%%%%%%%%%%%%%%%%%%%%%%%%%%%%%%%%%%%%%%%%%%%%%%%
\subsection*{Dynamical and elastic stability}

\begin{figure}
\centering
\includegraphics[width=0.6\columnwidth]{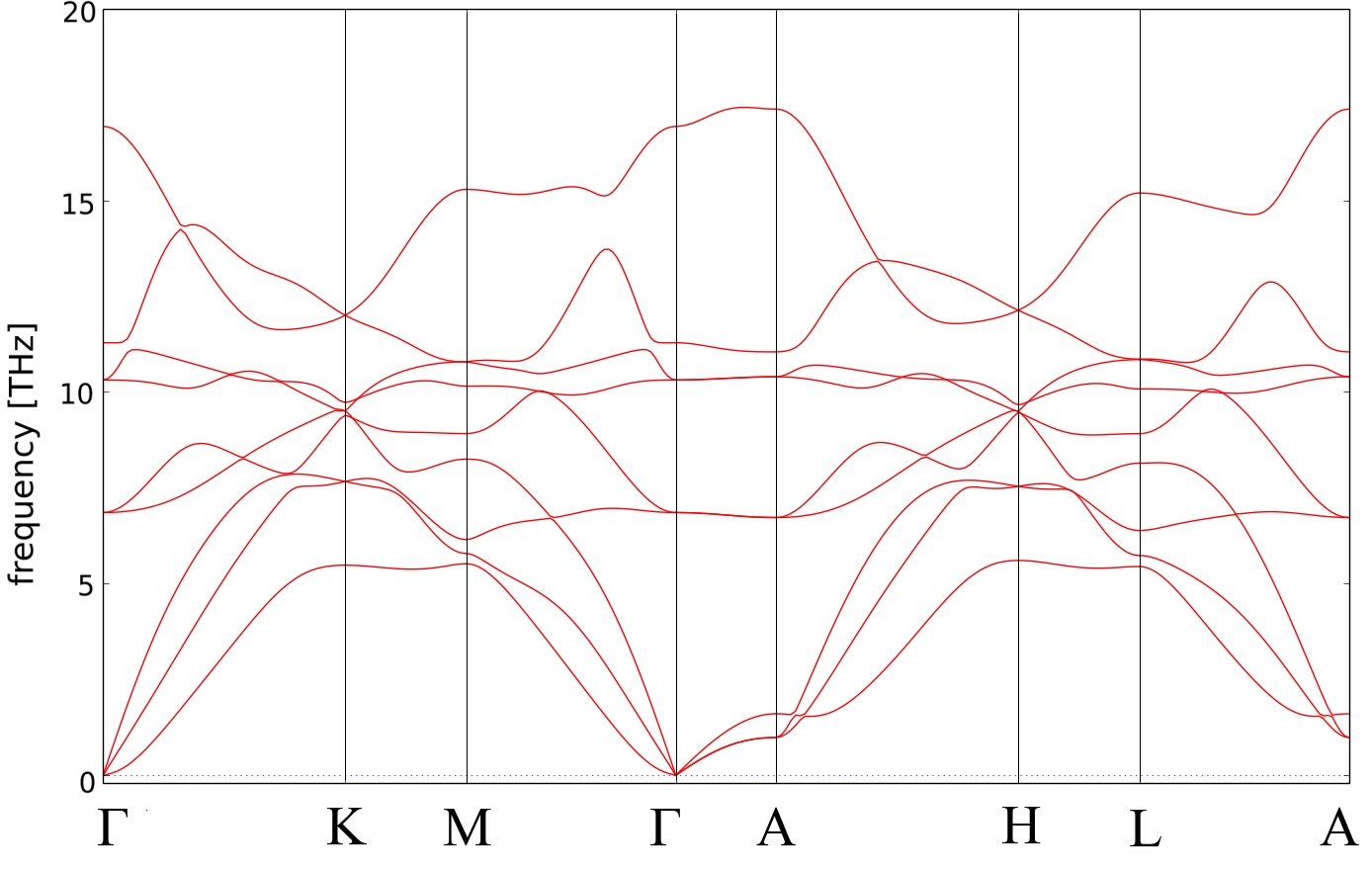}
\caption{Phonon dispersion curves for the $P\bar{3}m1$ structure at zero pressure along the selected high-symmetry lines calculated with the phonopy code \cite{phonopy}.
The structure shows no imaginary frequencies and might be therefore quenchable to ambient conditions.}
\label{phonons}
\end{figure}

By calculating phonon dispersions at 0, 10, 30, 60 GPa and at 100 GPa, we found that the $P\bar{3}m1$ phase
shows no imaginary frequencies from ambient pressure up to 1 Mbar, indicating its dynamical stability, at least at zero temperature.
In Fig. \ref{phonons}, phonon dispersion relations of the $P\bar{3}m1$ phase are shown for zero pressure. Modes for the two other
phases look, as one would expect, very similar.

\begin{table*}
\begin{tabular}{|l|l|c|l|l|}
\hline
phase & point group associated with space group & Laue group & independent elastic &  stability conditions \\
(space group) & and $C_{\alpha\beta}$ symmetry relations &  &   constants $C_{\alpha\beta}$ [GPa]   &   \\ \hline
$P\bar{3}m1$ & \hspace{2.6cm} $\bar{3}m$ & R~I ($\bar{3}m$) &            \hspace{1.6cm}  6                                &                                               \\ \hline
        &  $C_{22}$ = $C_{11}$          &                      &  $C_{11}$ = 192.8; $C_{33}$ = 25.4            &    $C_{11} > |C_{12}|$                              \\
        &  $C_{23}$ = $C_{13}$          &                      &  $C_{12}$ = 42.0; \, $C_{13}$ = 5.6             &  $2C_{13}^2 < C_{33}(C_{11}+C_{12})$                               \\
        &  $C_{24}$ = -$C_{14}$          &                      &  $C_{14}$ = 0.2                               &   $2C_{14}^2 < C_{44}(C_{11}-C_{12})$                              \\
        &  $C_{55}$ = $C_{44}$; $C_{66}$ = $\frac{1}{2}$($C_{11}$ - $C_{12}$)          &       &  $C_{44}$ = 7.7                               & $C_{44} > 0$  \\
        &  $C_{56}$ = $C_{14}$          &                      &                                               &                                 \\ \hline
$P6_3mc$ &  \hspace{2.5cm} $6mm$    &  H~I ($6/mmm$)   &          \hspace{1.6cm}    5                                &                                                    \\ \hline
         &  $C_{22}$ = $C_{11}$            &                      &  $C_{11}$ = 191.5; $C_{33}$ = 25.3            &       $C_{11} > |C_{12}|$                        \\
         &  $C_{23}$ = $C_{13}$          &                      &  $C_{12}$ = 41.2; \, $C_{13}$ = 5.2             &       $2C_{13}^2 < C_{33}(C_{11}+C_{12})$        \\
         &  $C_{55}$ = $C_{44}$; $C_{66}$ = $\frac{1}{2}$($C_{11}$ - $C_{12}$)          &     &  $C_{44}$ = 6.7    &  $C_{44} > 0$ \\
         &                                                                              &     &                    &  $C_{66} > 0$ \\ \hline
$R\bar{3}m$  & \hspace{2.6cm} $\bar{3}m$ & R~I ($\bar{3}m$) &       \hspace{1.6cm}       6                                &                                                    \\ \hline
        &            &                      &  $C_{11}$ = 190.9; $C_{33}$ = 22.4            &                                 \\
        &            &                      &  $C_{12}$ = 40.0; \, $C_{13}$ = 5.6             &                                 \\
        &            &                      &  $C_{14}$ = 0.1                               &                                \\
        &            &                      &  $C_{44}$ = 4.1                               &  \\ \hline
\end{tabular}
\label{elastic}
\caption{Elastic constants $C_{\alpha\beta}$ of the layered forms at zero pressure and temperature, symmetry relations and Born stability conditions.
Note that the Laue groups of $P\bar{3}m1$ and $R\bar{3}m$ space groups are the same.}
\end{table*}

Elastic constants $C_{\alpha\beta}$ (in Voigt notation) listed in Table \ref{elastic} were calculated at fixed volumes corresponding to zero pressure 
and at $T$ = 0 K. The Table also shows point groups of our investigated phases and their corresponding Laue groups as well as the respective symmetry relations between 
elastic constants and elastic stability conditions. For all structures there is about a factor of eight difference in the value of $C_{33}$ (corresponding to the direction 
perpendicular to planes) with respect to $C_{11}=C_{22}$. This indicates highly anisotropic mechanical properties as expected for layered systems.
All three investigated structures were found to be mechanically stable at zero pressure by fulfilling the necessary and sufficient Born stability conditions
for their corresponding Laue groups \cite{Mouhat-Coudert}, which come from the requirement of positive definiteness
of the $C_{\alpha\beta}$ matrix. Therefore, recovery of these phases in a metastable state at ambient pressure is at least conceptually possible. 

%%%%%%%%%%%%%%%%%%%%%%%%%%%%%%%%%%%%%%%%%%%%%%%%%%%%%%%%%%%%%%%%%%%%%%%%%%%%%%%%%%%%%%%%%%%%%%%%%%%%%%%%%%%%%%%%%%%%
\subsection*{Monolayer SiS$_2$}

\begin{figure}
\centering
\includegraphics[width=\columnwidth]{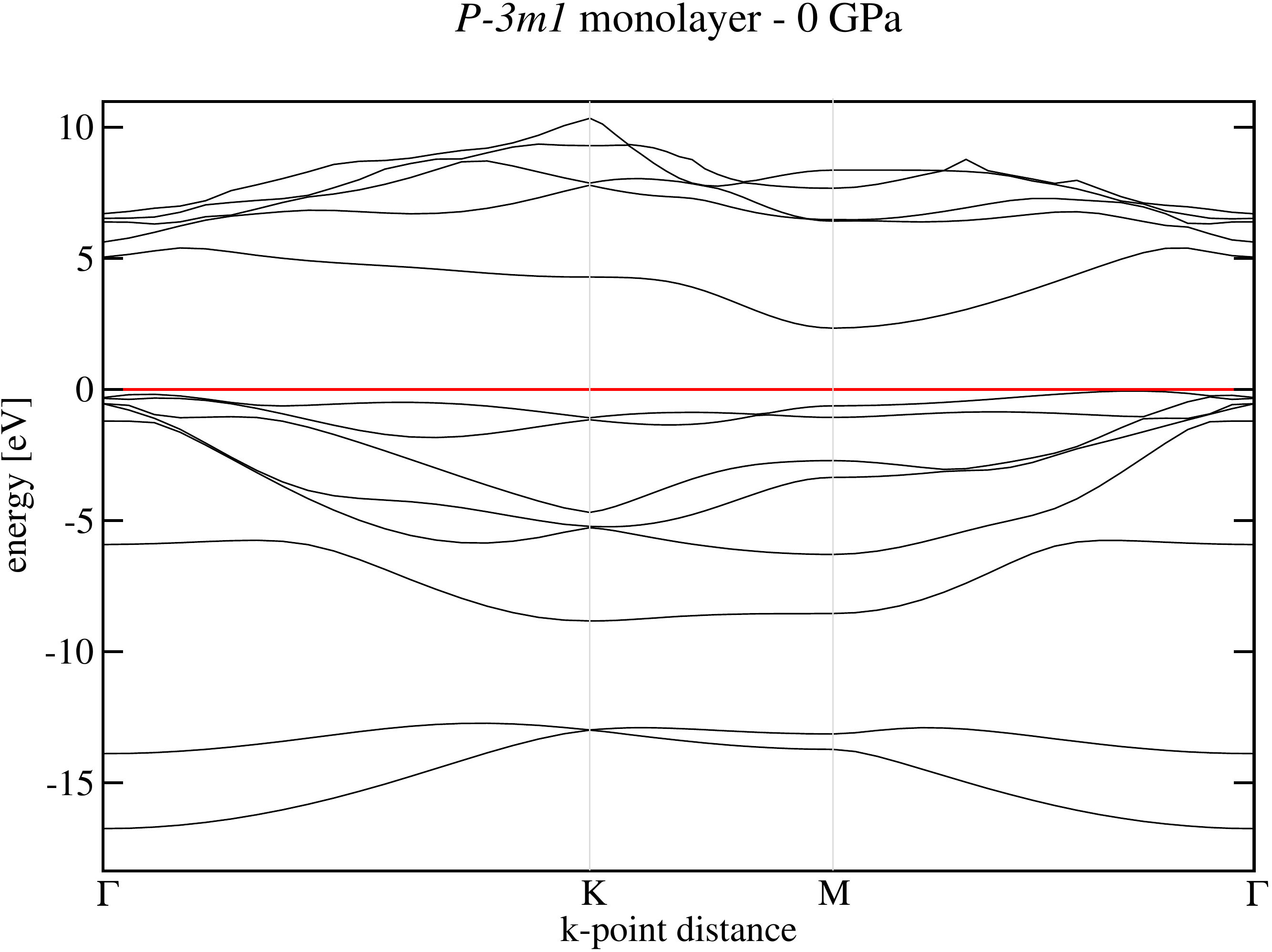}
\caption{Electronic band structure of a single octahedral monolayer calculated with HSE06 functional.}
\label{monolayer}
\end{figure}

The dynamical and elastic stability of the $P\bar{3}m1$ layered phase at all pressures down to zero has several consequences. First, it should be possible to shear
the layers relatively easily, as one does in graphite and in MoS$_2$. Again by similarity, one can hypothesize the possibility to exfoliate
a single octahedral SiS$_2$ monolayer from bulk $P\bar{3}m1$ and recover it at ambient conditions. To pursue this idea, we carried out a separate study 
of an SiS$_2$ monolayer. After geometrical optimization in a cell with 15 \AA \, vacuum between monolayers, we calculated its electronic structure 
along $k_z$=0 path - Fig. \ref{monolayer}. The band structure shows that the monolayer is again an indirect semiconductor with gap of about 2.4 eV.
In addition, we also calculated phonon dispersions of the monolayer and, as in the case of 3D layered octahedral phases, we found no imaginary modes indicating 
its dynamical and mechanical stability in isolation. Tetrahedral monolayer taken as one layer from HP1-SiS$_2$ has an indirect band gap larger than 3 eV.

%%%%%%%%%%%%%%%%%%%%%%%%%%%%%%%%%%%%%%%%%%%%%%%%%%%%%%%%%%%%%%%%%%%%%%%%%%%%%%%%%%%%%%%%%%%%%%%%%%%%%%%%%%%%%%%%%%%%
\subsection*{Molecular Dynamics of non-layered-to-layered structural transformation }

\begin{figure}
\centering
\includegraphics[width=\columnwidth]{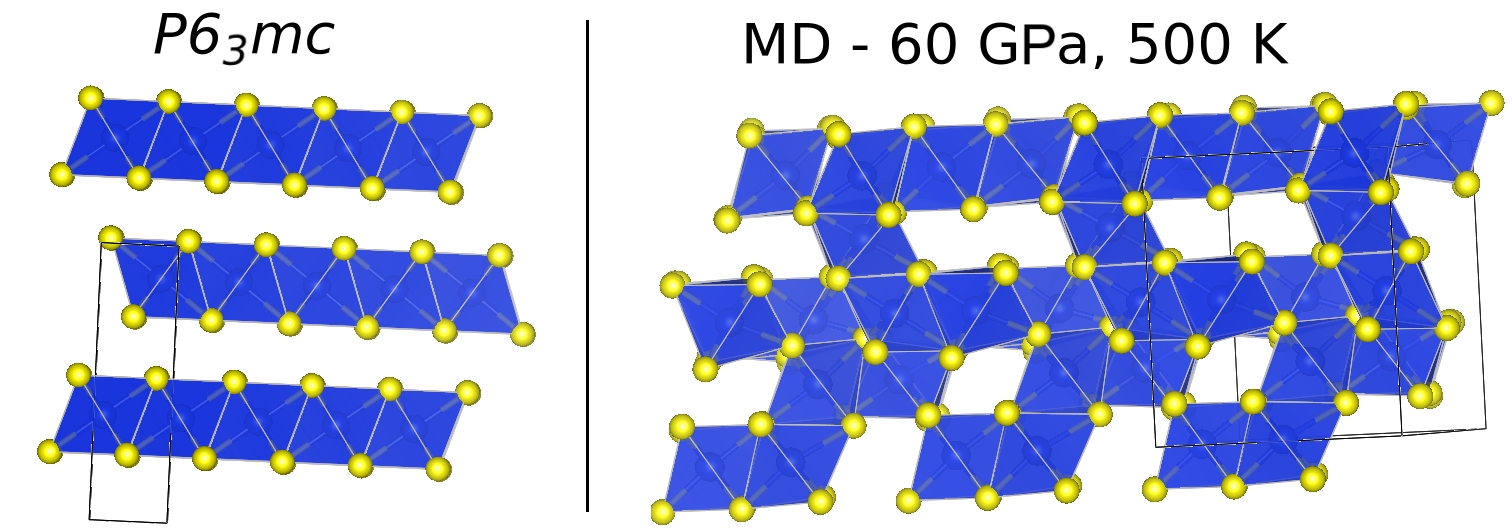}
\caption{Comparison of the ideal $P6_3mc$ structure and result of MD at 60 GPa and 500 K. Planes arising from irregularly connected octahedra
are well-visible from the dynamical simulations.}
\label{MD}
\end{figure}

As in all first-order transitions, the transformation from non-layered to layered structures will proceed by nucleation. 
While nucleation is in itself extremely slow and hard to simulate, much can still be learned about the local mechanism of transformation
by means of variable cell constant-pressure \textit{ab initio} molecular dynamics \cite{PR-2} which artificially
permits the transformation to take place uniformly and in extremely short times.

We therefore conducted $NPT$ $ab$ $initio$ MD simulations starting from the initial structure of HP3-SiS$_2$ for 48 and for 96 atoms 
at low pressure and room temperature and continuing by increasing pressure and temperature. In both sizes
we observed a chemical reorganization into fully octahedral forms - at the pressure of 60 GPa and temperature 600 K for
the 96-atomic system and at 50 GPa and 500 K for the 48-atomic system. The initial transformations led to irregular connectivity pattern 
of the SiS$_6$ units, which shared edges and corners and also contained unpaired S atoms. Upon further compression of the 48-atomic
system to 60 GPa at 500 K,  the octahedra shifted and rotated, transforming into an incompletely layered state with numerous defects.
Nevertheless, a strong resemblance to the $P6_3mc$ phase that contains two sheets per unit cell is visible from the polyhedral
view of $P6_3mc$ structure and MD result at 60 GPa - Fig. \ref{MD}.

MD calculations therefore confirm  the existence of octahedral SiS$_2$  with layered character at high pressures, even though
due to the limited time scales of $ab$ $initio$ MD simulations, proper defectless crystalline structures were not fully recovered.
An additional reason for incomplete transition observed in MD in this specific case might be the fact that the HP3 phase, which should
transform into $P\bar{3}m1$, is not layered, but fully three-dimensional. The transformation into octahedral geometry
starts at random places in the (simulation) sample and the emerging octahedral arrangement of silicon and sulfur atoms formed within the initial
chemical reaction then remains mostly unchanged in further evolution at the picosecond time scale. Only a substantial diffusion of certain
Si atoms could create proper two-dimensional sheets from irregularly connected octahedra, and therefore the original disordered polyhedral state remains
stable on the accessible $ab$ $initio$ time scale.

From the observed results, it can be predicted that the transformation from HP3 (or HP2) to $P\bar{3}m1$ phase possibly proceeds in two
stages - chemical and topological. The first, chemical stage is the transformation of distorted tetrahedra of HP3 into octahedra,
which lasted only about 2 ps in our simulations. In the second, topological stage, octahedra must shift and rotate in order to properly organize into
individual layers, but during this process some atoms must diffuse over distances that are larger than distances of second-nearest neighbors,
which in a solid material takes considerably longer time than the initial chemical reaction. 

These results and conclusions are important in two respects. First, they probably explain why HP3 may have been created and survived experimentally 
even though metastable with respect to layered $P\bar{3}m1$.  In turn, they support the possibility that layered $P\bar{3}m1$, once created,
might survive as a metastable phase once brought down to lower or even zero pressure.   

%%%%%%%%%%%%%%%%%%%%%%%%%%%%%%%%%%%%%%%%%%%%%%%%%%%%%%%%%%%%%%%%%%%%%%%%%%%%%%%%%%%%%%%%%%%%%%%%%%%%%%%%%%%%%%%%%%%%
\subsection*{Structural similarities with isoelectronic compounds}

\begin{figure*}
\centering
\includegraphics[width=2.0\columnwidth]{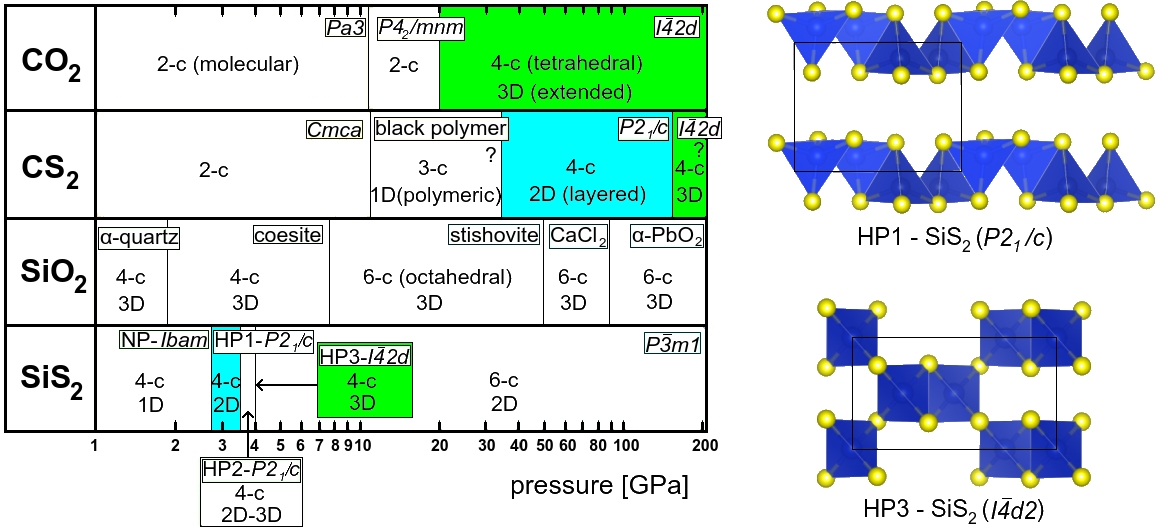}
\caption{Comparison of zero-temperature structures of CO$_2$ \cite{Santoro-Gorelli-CO2-review, Yoo-CO2-review-1, Iota-Yoo-Cynn, Yoo-1, Datchi-V, Santoro-3, Iota-Yoo, Yoo-2, Bonev, Sengupta-3, Tschauner, Togo, Serra, Sun},
CS$_2$ \cite{Yuan-Ding, Dias, Zarifi, Naghavi},
SiO$_2$ \cite{Tse-SiO2, Dove, Murakami, Prakapenka-SiO2, Dera, Nekrashevich-Gritsenko, cell-metadynamics-3, Hu-SiO2} and SiS$_2$ \cite{Silverman-Soulen, Prewitt-Young, Peters-Krebs, Guseva, Evers} showing a rather weak level of similarity.
The two structures $P2_1/c$ and $I\bar{4}2d$ that are common for some systems are highlighted in the diagram and shown for the case of SiS$_2$.
For CO$_2$, phase V is usually created during heating only at pressures over 50 GPa though it is predicted to be thermodynamically stable from 20 GPa \cite{Sengupta-3, Tschauner, Togo}.}
\label{phase_diagrams}
\end{figure*}

Finally, we should comment on structural similarities among the isovalent compounds CO$_2$, CS$_2$, SiO$_2$ and SiS$_2$. With hindsight, before the advent of efficient
crystal structure prediction techniques it was customary to argue based on assumption of similarity of the tetrahedral structures, in particular
to those of well-known SiO$_2$ \cite{Tse-SiO2, Dove, Murakami, Prakapenka-SiO2, Dera, Nekrashevich-Gritsenko, cell-metadynamics-3, Hu-SiO2}.
With the current knowledge of phase diagrams many such arguments no longer stand. In Fig. \ref{phase_diagrams}
we show the diagrams of the $T$=0 stable structures of the four compounds in a wide range of pressures up to 200 GPa. Restricting to these stable
structures there is in reality only a limited amount of similarity. The only tetrahedral structures that are stable in more than one compound are the monoclinic
$P2_1/c$ (whose existence in CS$_2$ is predicted, even not yet experimentally demonstrated) and tetragonal $I\bar{4}2d$ (if it is indeed
stable and not metastable in SiS$_2$ and stable in CS$_2$, which also remains to be experimentally demonstrated).
Considering also metastable structures, the only tetrahedral structure which seems to be universal at least topologically
is the $\beta$-cristobalite structure which is found in its ideal cubic form as a metastable phase of SiO$_2$ and in its distorted form $I\bar{4}2d$
stable in CO$_2$ and stable or metastable in SiS$_2$ (possibly stable in CS$_2$ too). Surprisingly, the lowest amount of structural similarity to other compounds is found
in SiO$_2$ while the highest is seen in SiS$_2$ and CS$_2$. In particular, the well-known quartz structure, which in its $\alpha$-form represents
the stable form of SiO$_2$ at ambient pressure, does not seem to be realized in stable form anywhere else; and the same is true for the coesite structure.
This comparison clearly shows that simple analogies are of limited value if used as heuristic guide in the family of binary compounds that we considered, and 
probably also elsewhere.

%%%%%%%%%%%%%%%%%%%%%%%%%%%%%%%%%%%%%%%%%%%%%%%%%%%%%%%%%%%%%%%%%%%%%%%%%%%%%%%%%%%%%%%%%%%%%%%%%%%%%%%%%%%%%%%%%%%%
\section*{Discussion}

In summary, we predicted by a highly reliable theoretical  protocol three new low-enthalpy high-pressure phases of SiS$_2$, namely $P\bar{3}m1$, $P6_3mc$ and $R\bar{3}m$
by means of $ab$ $initio$ structure searching method based on evolutionary algorithm. These three novel phases are all formed by SiS$_6$ octahedral
units forming separate sheets and differ only by the stacking pattern of the layers. All phases are predicted to become
more stable than the tetrahedral forms beyond 4 GPa at zero temperature and remain to be the three most stable structures up to at least 100 GPa,
out of all structures found. Upon increase of pressure SiS$_2$ therefore undergoes an interesting evolution as far as the dimensionality is concerned - tetrahedral structures
transform from 1D to 2D to 3D and after transition to octahedral structure it settles on a  2D layered character.
The associated density increase for tetrahedral-to-octahedral transformation is predicted to be around 20\% at 4 GPa showing a strong first-order nature of the transition.
The associated high barriers, and the large reorganization demanded by the transformation suggests the possibility to recover the new phases 
in metastable form even at ambient conditions.
Within the employed calculation scheme, the HP3 phase appears to be metastable at zero temperature. Enthalpies of the layered octahedral phases are
very similar at lower pressures, which indicates a possible presence of a layered structure with irregular stacking of sheets at experimental conditions. 
In constant-pressure molecular dynamics simulations, the spontaneous appearance of octahedral geometry starting from an initial tetrahedral one was also 
observed at 50 GPa, even though the resulting octahedral structure remained mostly disorganized due to short simulation times accessible in first-principles studies.
All three low-enthalpy octahedral polymorphs are semiconductors with indirect ~2 eV order band gaps at low pressures, values potentially interesting for layered compound applications.
They metallize around 40 GPa and remain relatively poor metals up to 100 GPa. The $P\bar{3}m1$ phase is dynamically stable from zero pressure
up to at least 1 Mbar from phonon calculations. Elastic stability of all phases was also confirmed at zero pressure and the calculated elastic constants indicate 
highly anisotropic mechanical properties typical for layered structures. Calculations also suggest that single octahedral layers could possibly be exfoliated 
from bulk $P\bar{3}m1$-SiS$_2$ once quenched to ambient conditions, as far as the monolayer is predicted to be dynamically stable once formed. Results presented 
in this study may be useful in order to compare the future experimental compression products with the ideal phases found in our structural search. On account of their
layered nature, the application of high-pressure shear may suffice to cause interlayer sliding, with transformations and the appearance of new polytype phases. 
Comparing the $T$=0 phase diagrams of selected group IV-VI AB$_2$ systems shows that structural similarities between these isovalent compounds
are in fact rather limited.

%%%%%%%%%%%%%%%%%%%%%%%%%%%%%%%%%%%%%%%%%%%%%%%%%%%%%%%%%%%%%%%%%%%%%%%%%%%%%%%%%%%%%%%%%%%%%%%%%%%%%%%%%%%%%%%%%%%%
\section*{Methods}

We employed the open-source evolutionary algorithm package for crystal structure prediction XtalOpt \cite{XtalOpt} and carried out searches
at 10, 30, 60 and at 100 GPa with 6 and 12 atom cells (2 and 4 structural units) for each pressure, enabling generation of at least 1000 structures in each case.
All underlying \textit{ab initio} electronic calculations were performed with the density functional theory (DFT) VASP 5.3 and 5.4 codes \cite{VASP-1, VASP-2}
employing projector augmented-wave pseudopotentials \cite{Blochl-PAW, VASP-PAW} and Perdew-Burke-Ernzerhof (PBE) parametrization of GGA exchange-correlation functional \cite{PBE}. 
Pressures above 100 GPa could not be, and were not, addressed with these pseudopotentials.
The randomly generated initial structures of the evolutionary search and those created after various mutations
were initially optimized in static cells (where only ions were relaxed) and within coarse $k$-point grids. Thereafter, structures were
relaxed in fixed-volume geometry (ions and cell shapes were relaxed) with a finer $k$-points mesh. Finally, full ionic and cell degrees
of freedom relaxations were carried out with finest $k$-point sampling in order to refine the final structures. During this progressive
structural optimization, the energy cutoff was also gradually increased from 340 to 400 eV.

In order to describe the dispersion forces that must be included for the obtained layered phases, especially in the low-pressure region, we employed
the parameter-free Tkatchenko-Scheffler approach DFT-TS \cite{Tkatchenko-Scheffler}, in which dispersion coefficients and damping functions of the original Grimme's
method \cite{Grimme-1, Grimme-2} are charge-density dependent. The dispersion forces are therefore described adaptively accounting for changing electronic
structure upon compression. To investigate electronic structure of octahedral SiS$_6$, we employed the HSE06 hybrid functional \cite{HSE06}, which tends to provide
more reliable results for the band gap estimate and electronic structure of solids compared to PBE.

The second-order elastic stiffness constants $C_{\alpha\beta}$ were calculated using VASP by a two-steps procedure, in which the stress tensor
is determined for both distorted supercells without relaxing the ions and for undistorted supercells with separately shifted ions \cite{Wu-Vanderbilt-Hamann}.
For each phase, we used 21 different values of strain - from 0.005 to 0.025 in 0.001 intervals and from these, only results that were quite similar
(formed plateaus on $C_{\alpha\beta}$ vs. strain dependences) were used to obtain final average - this was from strain values 0.015 to 0.02 for $P\bar{3}m1$ phase,
between 0.011-0.021 for the case of $P6_3mc$ and for 0.013-0.016 interval for $R\bar{3}m$.
From the final elastic tensor of each phase, individual $C_{\alpha\beta}$ constants were extracted by requirements of symmetry according
to each of the phases' Laue groups \cite{Wallace}. For example, if, say, $C_{11}$ should by symmetry be equal to $C_{22}$, then $C_{11}$ was
taken to be ($C_{11}$ + $C_{22}$)/2, and so on. Elastic constants for all phases were calculated at zero temperature at volumes corresponding
to zero pressure and with the use of Grimme's van der Waals correction (DFT-D2) \cite{Grimme-1, Grimme-2}.

Phonon dispersion curves were calculated in the harmonic approximation by the supercell method using the phonopy code \cite{phonopy}.
A $6\times 6\times 4$ supercell (432 atoms) for the $P\bar{3}m1$ phase, a $6\times 6\times 2$ supercell (432 atoms) for the $P6_3mc$ phase
and a $6\times 6\times 2$ supercell (648 atoms) for $R\bar{3}m$ phase were used for calculations of force constant matrices with phonopy.
Correctness of the dispersion curves was also tested in some important cases by independent calculations using density functional perturbation
theory as employed in the Quantum Espresso package \cite{QEspresso}.

MD simulations were performed in the constant pressure-temperature \textit{NPT} ensemble with the use of Parrinello-Rahman barostat \cite{PR-2} and Langevin stochastic thermostat \cite{Allen-Tildesley}.
The simulation samples contained 48 and 96 atoms and the supercells were generated as $4 \times 4 \times 4$ and $4 \times 4 \times 8$ unit-cells of the
HP3-SiS$_2$ structure. The time step for MD was set to 2 fs.

%%%%%%%%%%%%%%%%%%%%%%%%%%%%%%%%%%%%%%%%%%%%%%%%%%%%%%%%%%%%%%%%%%%%%%%%%%%%%%%%%%%%%%%%%%%%%%%%%%%%%%%%%%%%%%%%%%%%
%\bibliography{/home/dusanko/Desktop/Projects/references}

%%%%%%%%%%%%%%%%%%%%%%%%%%%%%%%%%%%%%%%%%%%%%%%%%%%%%%%%%%%%%%%%%%%%%%%%%%%%%%%%%%%%%%%%%%%%%%%%%%%%%%%%%%%%%%%%%%%%
\section*{Acknowledgements}

This work was supported by the Slovak Research and Development Agency under Contract APVV-15-0496,
by the VEGA project No.~1-0904-15 and by the project implementation 26220220004 within the Research \& Development Operational Programme funded by the ERDF.
Part of the calculations were performed in the Computing Centre of the Slovak Academy of Sciences using the supercomputing
infrastructure acquired in project ITMS 26230120002 and 26210120002 (Slovak infrastructure for high-performance computing)
supported by the Research \& Development Operational Programme funded by the ERDF. Work in Trieste was carried out under 
ERC Grant 320796 MODPHYSFRICT. EU COST Action MP1303 is also gratefully acknowledged.

%%%%%%%%%%%%%%%%%%%%%%%%%%%%%%%%%%%%%%%%%%%%%%%%%%%%%%%%%%%%%%%%%%%%%%%%%%%%%%%%%%%%%%%%%%%%%%%%%%%%%%%%%%%%%%%%%%%%
\section*{Author contributions statement}
R.M. and E.T. designed research. D.P. performed research. D.P., R.M. and E.T. analyzed the data. D.P., R.M. and E.T. wrote the paper.

\end{document}